# State Ensembles and Quantum Entropy

Subhash Kak

ABSTRACT
This paper considers quantum communication involving an ensemble of states. Apart from the von Neumann entropy, it considers other measures one of which may be useful in obtaining information about an unknown pure state and another that may be useful in quantum games. It is shown that under certain conditions in a two-party quantum game, the receiver of the states can increase the entropy by adding another pure state.

1. INTRODUCTION

Thermodynamic entropy measures the number of ways in which the microstates of the system may be arranged, and it is equivalent to informational (Shannon) entropy which is simply the amount of information needed to specify the state of the system. It may also be understood as a measure of disorder.

The problem of information in a quantum system is not of a straightforward generalization of the corresponding classical concept. The two situations are not completely symmetric since in both the classical and the quantum case the observer of information is a classical system. Furthermore, both the sending and the receiving agents of information must make choices, which means that in the quantum case the behavior of the agents cannot be unitary.

Von Neumann derived his measure of quantum entropy [1],[2], using arguments similar to that of Boltzmann in his earlier derivation of thermodynamic entropy, by counting the different microstates associated with the system. The von Neumann measure of a pure state, even if it is an unknown state, is zero. This latter property is unsatisfactory from a communications perspective because if the sender is sending many copies of an unknown pure state to the receiver, it should, in principle, communicate information. In the classical case, the contents of an entire book may be converted to a single signal by converting the binary file of the book into a number by putting a "dot" before it to make it into a decimal number [3], and if a large number of copies of this signal are sent, the receiver could, in principle, determine its value even in the presence of noise. Likewise, many copies of an unknown pure quantum state can be used to obtain information about the state.

Although Shannon entropy is generally used in a communication situation, it is not the only measure of information; other measures with different domains of applicability include those of Fisher [4], Renyi [5], and Tsallis[6]. More recently, Fisher information together with the Bohm's quantum potential has been proposed for quantum information [7],[8]. Although the Fisher measure is useful statistically, it is problematic when considered in a communications setting. Thus this information in a Bernoulli process (say fair coin tosses) of $n$ trials is $4n$, whereas using our intuition about information it should only be n bits, one for each trial.

This paper considers a quantum communication situation in which the sender sends out an ensemble of states to the receiver. It reviews three measures of quantum entropy that may be considered for this situation. These include the von Neumann measure and two more that are inspired by the classical communication entropy of Shannon. Since different sets of mixed and pure states lead to the same density matrix, the actual sets used by the state preparer cannot be determined. We consider the extent that the entropy in a given state can be changed by adding another state to the ensemble with a predefined probability.

2. INFORMATION IN CLASSICAL AND QUANTUM SETTINGS

In communication theory, information is understood in terms of the yes-no type of decisions that can be made by the receiver upon receipt of the message. It is based on the difference between the a priori and the a posteriori probabilities associated with the messages which implies that their statistical behavior is known.

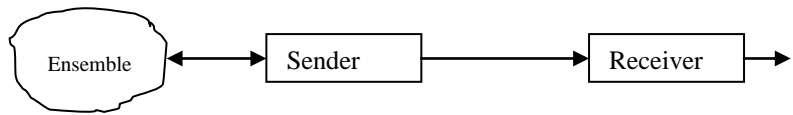

Figure 1. The information exchange setting

In communication, decision-making occurs both at the source and the receiver. The source must pick a specific message out of the ensemble at each sending instant and the receiver must estimate what was sent since, in general, he operates in a situation of uncertainty due to noise or incomplete information. The decision making must be non-unitary in the quantum case, since it involves a reduction of the a priori probability.

The measure of information is not limited to the probabilities of single messages and it also applies to strings of bits. As more data accumulates, the probabilities between signals (or messages) that are ever more separated in time are revealed. These probabilities can only make sense to the analyzer of the strings, who is the experimenter (or a computer may be programmed to make the estimation). In other words, the consideration of probabilities associated with an experimental scheme that involves senders and receivers posits a mind.

Information is a top-down property and, therefore, to speak of conservation of information [9]-[14] is very different from the conservation of energy, momentum, charge, or other similar quantities that have been traditionally examined in theoretical physics. The desire to see conservation of information as a universal principle parallels that of seeing unitarity as the fundamental transformational mechanism. Conservation of information does not appear to have meaning in the absence of conscious agents.

Questions of how the information may be unavailable due to noise [15] or deliberately concealed [16] are also relevant, as is the matter of the characteristics of the physical system processing this information [17].

In the classical communication problem the mutual information between two random variables X (sender's output) and Y (receiver's input) is given by *I(X;Y) = H(X) – H(X/Y),* where *H(X)* and *H(X/Y)* are the entropy values of X before and after the communication has been received. In the quantum communication situation, the Holevo bound [18],[19] tells us the amount of information that can be known about a quantum state. Assuming that the sender pick out of a set of states {ρ$_1$, ρ$_2$, ..., ρ$_n$} with probability {p$_1$, p$_2$, ..., p$_n$}, for any POVM measurement performed on $\rho = \sum_i p_i \rho_i$, the amount of information about the variable X knowing the outcome Y of the measurement is bounded as follows:

$$I(X,Y) \leq S_n(\rho) - \sum_i p_i S_n(\rho_i) \qquad (1)$$

where $S_n(\rho_i)$ is the entropy associated with $\rho_i$. If we fit our problem to the Holevo model, the contribution of the pure state to the information will be zero.

3. THREE MEASURES OF ENTROPY

The three measure considered here are the von Neumann, the informational, and the composite informational measure. Every density operator may be viewed as a mixture of pure states

$$\rho = \sum_i \lambda_i |\varphi_i\rangle\langle\varphi_i| \qquad (2)$$

where $\lambda_i$ are the eigenvalues and $|\varphi_i\rangle$ are the eigenvectors of ρ.

*Von Neumann entropy, S$_n$(ρ)*

Quantum information is generally measured by the von Neumann entropy [1], $S_n(\rho) = -tr(\rho \log \rho)$, where ρ is the density operator associated with the state ( "log" in the expression to base 2 implies that the units are *bits*). It plays a central role in many areas of physics [2],[7],[20] and it is important in the consideration of entanglement. In the general case, the ensemble of states has both mixed and pure components. The entropy may be written as

$$S_n(\rho) = -\sum_i \lambda_i \log \lambda_i \qquad (3)$$

where $\lambda_i$ are the eigenvalues of ρ. Thus the measurements along the reference bases may be associated with probability values $\lambda_i$ in analogy with the classical entropy expression of $-\sum_i p_i \log p_i$, where the *i*th outcome, out of a given set, has probability *p$_i$*.

For a bipartite system AB, where we consider the quantum states in pairs, we see the subadditive property:

$$S_n(\rho_{AB}) \leq S_n(\rho_A) + S_n(\rho_B) \tag{4}$$

with equality if the two systems are uncorrelated and $\rho_{AB} = \rho_A \otimes \rho_B$. This is of course similar to the subadditive property of classical information: $H(X,Y) \leq H(X) + H(Y)$.

Now consider the problem of state preparation. Let the state be drawn randomly from the ensemble $\{|\varphi_i\rangle, p_i\}$, $i=1,2,3$, where $|\varphi_1\rangle$ and $|\varphi_2\rangle$ are orthogonal and they form a distinguished subset of this ensemble and $|\varphi_3\rangle$ is a superposition of $|\varphi_1\rangle$ and $|\varphi_2\rangle$. We wish to see the influence on entropy of adding a new state to the ensemble.

In the somewhat parallel classical situation, the sender sends either a 0 or a 1, each of which carries one bit of information, or an intermediate value $m$, which carries no information. If the receiver should receive a string 01m10m1.., he will know that it needs to be changed to 01101… since information is additive in preparation as well. But this is not true of the quantum case. The pure state $|\varphi_3\rangle$ that carries new information influences the information carried by the other two states. Let's illustrate it with an example where
$$\rho = \begin{bmatrix} 0.7 & 0.2 \\ 0.2 & 0.3 \end{bmatrix}.$$

The eigenvalues of $\rho$ are 0.783 and 0.217 and, therefore, the von Neumann entropy $S_n(\rho)$ for this case is equal to

$$S_n(\rho) = -0.783 \log_2 .783 - 0.217 \log_2 .217$$
$$= 0.755 \text{ bits.}$$

This density matrix may be seen as a sum of the two components $\rho_1$ from $|\varphi_1\rangle$ and $|\varphi_2\rangle$ and $\rho_2$ from $|\varphi_3\rangle$ as follows:

$$\rho = 0.6 \times \begin{bmatrix} 0.833 & 0 \\ 0 & 0.167 \end{bmatrix} + 0.4 \times \begin{bmatrix} 0.5 & 0.5 \\ 0.5 & 0.5 \end{bmatrix} = 0.6 \times \rho_1 + 0.4 \times \rho_2$$

where $S_n(\rho_1) = -0.833 \log_2 .833 - 0.167 \log_2 .167 = 0.650$, and $S_n(\rho_2) = 0$, for $\rho_2$ represents a pure state. Thus,

$$S_n(\rho) \neq S_n(\rho_1) + S_n(\rho_2)$$

This means that state preparation does not satisfy additivity under the assumptions that were described.

*Quantum informational entropy*, $S_i(\rho)$

The measure of quantum informational entropy, $S_i(\rho)$ [21] is taken to equal:

$$S_i(\rho) = -\sum_i \rho_{ii} \log \rho_{ii} \qquad (5)$$

This measure does not depend on the off-diagonal terms of the density matrix and this may be considered its limitation. $S_i(\rho)$ may be conceived as a particular sum of the von Neumann entropy and the entropy associated with the unknown pure state where the contribution of the pure state is implicit for $S_i(\rho)$ overbounds the measure $S_n(\rho)$. It assumes the least about the structure of the density matrix and, therefore, it seeks to obtain the most information that is possible from a measurement. One might speculate that informational measure of entropy has connections with computational complexity [22].

*Composite informational entropy*, $S_{ci}(\rho)$

We define $S_{ci}(\rho)$, the composite informational entropy, as the sum of its mixed and pure components in an explicit manner:

$$S_{ci}(\rho) = -p_0 \sum_i \lambda_i \log \lambda_i + \sum_i p_i S_p(\phi_i) \qquad (6)$$

where $S_p(\phi)$ represents the informational entropy of the pure state $|\phi\rangle = \sum_k c_k |a_k\rangle$:

$$S_p(\phi) = -\sum_k |c_k|^2 \log |c_k|^2 \qquad (7)$$

For a qubit, the expression (6) simplifies to:

$$S_{ci}(\rho) = p_0 S_n(\rho) + p_1 S_p(\phi_i) \qquad (8)$$

For a source whose output is a mixed state with off-diagonal terms that are zero, the three entropy values $S_n(\rho), S_i(\rho),$ and $S_{ci}(\rho)$ are identical. The less assumed about the process, the more the entropy. For example, when off-diagonal terms are added to the two-dimensional density matrix, the eigenvalues become more different from each other implying reduced entropy. In other words, based on the amount of knowledge that is assumed, one may conclude:

**Theorem**. $S_n(\rho) \leq S_{ci}(\rho) \leq S_i(\rho)$.

4. ENSEMBLE OF MIXED AND PURE STATES

Assume that an ensemble of qubits is prepared by a source, S. We assume that S picks up the states $|0\rangle$, $|1\rangle$, and $u|0\rangle + v|1\rangle$ with probabilities $p_0, p_1,$ and $p_2$, respectively, where $u$ and $v$ will be taken to be real for simplicity. The density matrix is therefore:

$$\rho = p_0 \begin{bmatrix} 1 & 0 \\ 0 & 0 \end{bmatrix} + p_1 \begin{bmatrix} 0 & 0 \\ 0 & 1 \end{bmatrix} + p_2 \begin{bmatrix} u^2 & uv \\ uv & v^2 \end{bmatrix} \quad (9)$$

This may be rewritten as:

$$\rho = \begin{bmatrix} p_0 + p_2 u^2 & p_2 uv \\ p_2 uv & p_1 + p_2 v^2 \end{bmatrix}$$

The eigenvalues of this matrix are:

$$\lambda_1 = \frac{1}{2}(1 + \sqrt{1 - 4p_0 p_1 - 4p_0 p_2 v^2 - 4p_1 p_2 u^2})$$
$$\lambda_2 = \frac{1}{2}(1 - \sqrt{1 - 4p_0 p_1 - 4p_0 p_2 v^2 - 4p_1 p_2 u^2}) \quad (10)$$

which shows that the von Neumann entropy $S_n(\rho) = -\lambda_1 \log \lambda_1 - \lambda_2 \log \lambda_2$ will consist of terms that are pairwise products of the underlying probabilities.

Its informational entropy, $S_i(\rho)$, is simply

$$S_i(\rho) = -(p_0 + p_2 u^2) \log(p_0 + p_2 u^2) - (p_1 + p_2 v^2) \log(p_1 + p_2 v^2) \quad (11)$$

The density matrix may be rewritten as:

$$\rho = (p_0 + p_1) \begin{bmatrix} \dfrac{p_0}{p_0 + p_1} & 0 \\ 0 & \dfrac{p_1}{p_0 + p_1} \end{bmatrix} + p_2 \begin{bmatrix} u^2 & uv \\ uv & v^2 \end{bmatrix} \quad (12)$$

The first part is the mixed state with probability of $p_0+p_1$ and the second part is the pure state with the probability of $p_2$. That this is not a canonical representation as clear from the fact that (7) may be decomposed in a variety of ways. For example,

$$\rho = \begin{bmatrix} 0.592 & 0.144 \\ 0.144 & 0.408 \end{bmatrix} = 0.4 \times \begin{bmatrix} 1 & 0 \\ 0 & 0 \end{bmatrix} + 0.3 \times \begin{bmatrix} 0 & 0 \\ 0 & 1 \end{bmatrix} + 0.3 \times \begin{bmatrix} 0.64 & 0.48 \\ 0.48 & 0.36 \end{bmatrix}$$
$$= 0.2512 \times \begin{bmatrix} 1 & 0 \\ 0 & 0 \end{bmatrix} + 0.3488 \times \begin{bmatrix} 0 & 0 \\ 0 & 1 \end{bmatrix} + 0.4 \times \begin{bmatrix} 0.847 & 0.36 \\ 0.36 & 0.153 \end{bmatrix}$$

The entropy may be written as:

$$S_{ci}(\rho) = -p_0 \log p_0 - p_1 \log p_1 + (p_0 + p_1) \log(p_0 + p_1) - p_2 u^2 \log u^2 - p_2 v^2 \log v^2$$

(13)

In the absence of the pure component, the entropy was $-p_0 \log p_0 - p_1 \log p_1$ where $p_0 + p_1 = 1$ and $p_2 = 0$.

**Special case, $p_0 = p_1$.** To investigate the relationship between the different entropy measures, we take the special case of $p_0 = p_1$, which maximizes the entropy of the mixed component state. In this case, the state is given by the density operator

$$\rho = \begin{bmatrix} 0.5 & a \\ a & 0.5 \end{bmatrix} \tag{14}$$

where $a$ varies from 0 to 0.5.

The eigenvalues for this example are $\lambda_1 = 0.5 + a$ and $\lambda_2 = 0.5 - a$. The von Neumann entropy is therefore

$$S_n(\rho) = -(0.5+a)\log(0.5+a) - (0.5-a)\log(0.5-a)$$

The informational entropy of this remains constant at 1, whereas the von Neumann entropy for this varies from 1 to 0 as $a$ ranges from 0 to 0.5 as shown in Figure 2. The density matrix of equation (14) may be represented as a sum of mixed and pure states as shown below:

$$\rho = (1-2a)\begin{bmatrix} 0.5 & 0 \\ 0 & 0.5 \end{bmatrix} + 2a\begin{bmatrix} 0.5 & 0.5 \\ 0.5 & 0.5 \end{bmatrix} \tag{15}$$

The second term on the right is the pure component and its share of the entropy will be one bit times its probability, that is $2a$ bits. This is shown in Table 1. We note that $S_i(\rho)$ and $S_{ci}(\rho)$ are identical for this example.

Table 1. Entropy values for state given by equation (14), $S_i(\rho) = S_{ci}(\rho)$

| $a$ | 0 | 0.05 | 0.10 | 0.15 | 0.20 | 0.25 | 0.30 | 0.35 | 0.40 | 0.45 | 0.50 |
|---|---|---|---|---|---|---|---|---|---|---|---|
| $S_i(\rho)$ | 1 | 1 | 1 | 1 | 1 | 1 | 1 | 1 | 1 | 1 | 1 |
| $S_p(\rho)$ | 0 | 0.1 | 0.2 | 0.3 | 0.4 | 0.5 | 0.6 | 0.7 | 0.8 | 0.9 | 1.0 |
| $S_n(\rho)$ | 1 | 0.993 | 0.971 | 0.934 | 0.881 | 0.811 | 0.722 | 0.610 | 0.469 | 0.286 | 0 |

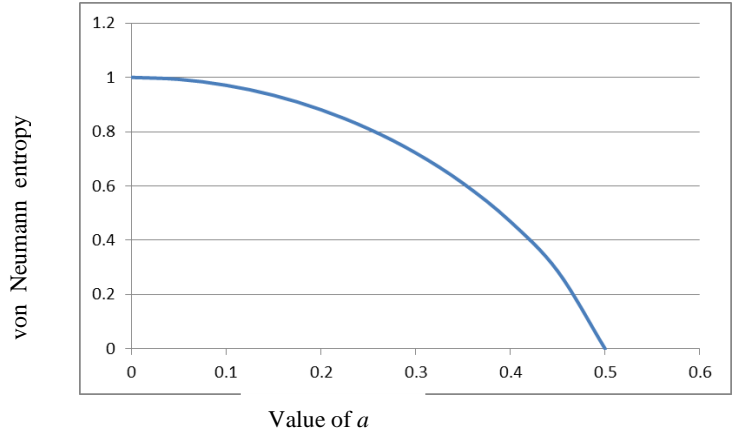

Figure 2. Change in the entropy with respect to *a*

Note that

$$S_i(\rho) \leq S_{pure}(\rho) + S_n(\rho) \tag{16}$$

The excess of the sum of $S_{pure}(\rho)$ and $S_n(\rho)$ over the value of informational entropy in the first row is maximum for the *a* value of 0.30.

**General case, $p_0 \neq p_1$.** Now we go back to the case that was originally mentioned in equation (8) with unequal probabilities $p_0$ and $p_1$ which correspond to unequal x and y:

$$\rho = \begin{bmatrix} x & a \\ a & y \end{bmatrix} \tag{17}$$

It may be represented as a sum of mixed and pure states in multiple ways. The general form is:

$$\rho = (x+y)\begin{bmatrix} \dfrac{x - p_2 u^2}{x+y} & 0 \\ 0 & \dfrac{y - p_2 v^2}{x+y} \end{bmatrix} + p_2 \begin{bmatrix} u^2 & uv \\ uv & v^2 \end{bmatrix} \tag{18}$$

where $p_2 uv = a$. This may be rewritten as:

$$\rho = (1 - p_2)\begin{bmatrix} \dfrac{x - p_2 u^2}{1 - p_2} & 0 \\ 0 & \dfrac{y - p_2 v^2}{1 - p_2} \end{bmatrix} + p_2 \begin{bmatrix} u^2 & uv \\ uv & v^2 \end{bmatrix} \tag{19}$$

where $p_2 uv = a$. As a special case, let $p_2 = 2a$, and $u^2 = v^2 = 0.5$. We get:

$$\rho = (1-2a)\begin{bmatrix} \dfrac{x-a}{1-2a} & 0 \\ 0 & \dfrac{y-a}{1-2a} \end{bmatrix} + 2a\begin{bmatrix} 0.5 & 0.5 \\ 0.5 & 0.5 \end{bmatrix} \qquad (20)$$

where both $x$ and $y$ are greater than $a$.

The total entropy for this case (11) may be rewritten as:

$$S_{ci}(\rho) = -(x-a)\log(x-a) - (y-a)\log(y-a) + (1-2a)\log(1-2a) + 2a \qquad (21)$$

As we see in Figure 3, the total entropy goes down with the change in value from $x = 0.5$ as well as increasing $a$.

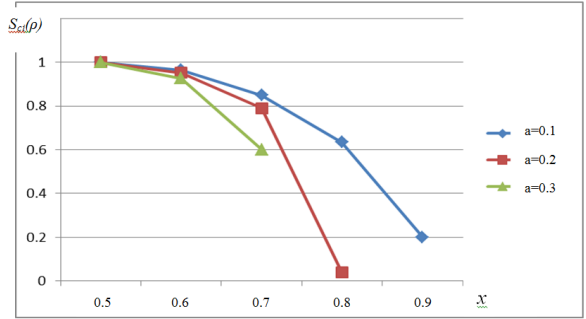

Figure 3. Change in the entropy with respect to $x$ and $a$

5. ENTROPY TRANFORMING SYSTEM

We consider the system given in Figure 4 in which B tries to transform the entropy of the states reaching him.

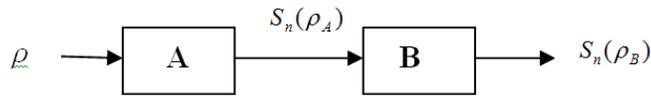

Figure 4. An entropy transforming circuit

Let A be preparing specific quantum states for transmission. B, intercepting these states, modifies them by adding further states from an available set. Our objective is to determine how much of a change of entropy can be achieved. Such a situation exists in certain models of quantum cryptography [16].

We assume that A prepares $\rho_A = \begin{bmatrix} \lambda & 0 \\ 0 & 1-\lambda \end{bmatrix}$ by choosing $|0\rangle$ and $|1\rangle$ with probabilities $\lambda$ and $1-\lambda$. B estimates the density matrix and then creates a new state by choosing this state with probability 0.5 and another pure state $\begin{bmatrix} \sqrt{(1-\lambda)} \\ \sqrt{\lambda} \end{bmatrix}$ also with probability 0.5. The density function of the state created by B will be:

$$\rho_B = \begin{bmatrix} 0.5 & \sqrt{\lambda(1-\lambda)}/2 \\ \sqrt{\lambda(1-\lambda)}/2 & 0.5 \end{bmatrix} \qquad (22)$$

The interesting question to ask is: For what values of λ does the value of $S_n(\rho_B)$ exceed $S_n(\rho_A)$? Th eigenvalues of (22) are $(.5+.5\sqrt{\lambda(1-\lambda)})$ and $(.5-.5\sqrt{\lambda(1-\lambda)})$. Therefore, our question may be answered by solving for:

$$-(.5+.5\sqrt{\lambda(1-\lambda)})\log_2(.5+.5\sqrt{\lambda(1-\lambda)}) - (.5-.5\sqrt{\lambda(1-\lambda)})\log_2(.5-.5\sqrt{\lambda(1-\lambda)}) \geq \\ -\lambda\log_2\lambda - (1-\lambda)\log_2(1-\lambda) \qquad (23)$$

Solving, we find that the range of λ for which this is true is (0,0.2805) and (0.7195,1.0) as shown in Figure 5.

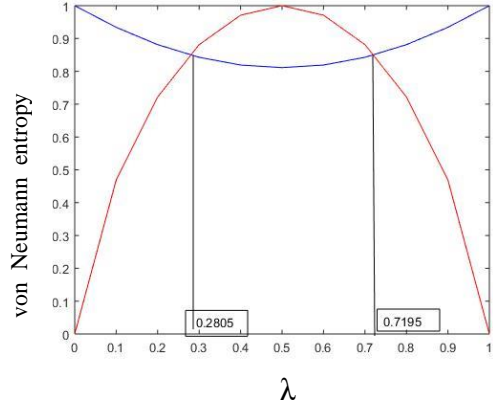

Figure 5. An entropy transforming system
(red line, original entropy; blue line, transformed entropy)

Although the density matrix can be decomposed in a variety of ways, the nature of the instruments available for measurement will constrain this decomposition. Indeed, one can imagine that the qubits are produced in a lab and mixed up with the appropriate probabilities they are sent out to the experimenter in the other room to estimate the amount of information in them as has also been investigated in other studies[23],[24]. This scenario may also be relevant in certain quantum games.

6. DISCUSSION

This paper reviews reasons why we may need more than one measure for quantum entropy. It shows that even subsets of ensembles that carry no information have an influence on the von Neumann entropy measure in a manner somewhat reminiscent of interaction-free measurement [25],[26]. If science is the game that we play with Nature to find an unknown pure state in some suitable configuration space, then this information (like $S_i(\rho)$) is basis dependent that is obtained by repeated measurements on the ensemble and our discovery of this preferred basis state is the culmination point of science.

The paper considered the problem of the decomposition of the single-qubit density operator as a probabilistic sum of pure and mixed component states and although this decomposition can be done in a variety of ways, it may in practice be constrained by the nature of the measurement apparatus available to the measuring agent as may be true in a quantum game. The measure for quantum informational entropy is for communication situations where an unknown pure state can communicate information. It is shown that under certain conditions in a two-party quantum game, the receiver of the states can increase the entropy by adding another pure state.

**Acknowledgements.** I am thankful to Kam Wai C. Chan for his comments on an earlier version of the paper. I am also thankful to the National Science Foundation for its support through grant #1117068.

7. REFERENCES
1. Von Neumann, J. Mathematische Grundlagen der Quantenmechanik, Springer, Berlin (1932)
2. Petz, D. Entropy, von Neumann and the von Neumann entropy. In John von Neumann and the Foundations of Quantum Physics, eds. M. Redei and M. Stoltzner, Kluwer (2001)
3. Kak, S. and Chatterjee, A. On decimal sequences. IEEE Trans. on Information Theory IT-27: 647 – 652 (1981)
4. Savage, L. J. On rereading R. A. Fisher. Annals of Statistics 4 (3): 441–500 (1976)
5. Rényi, A. On measures of information and entropy. Proceedings of the fourth Berkeley Symposium on Mathematics, Statistics and Probability 1960. 547–561 (1961)
6. Tsallis, C. Possible generalization of Boltzmann-Gibbs statistics". Journal of Statistical Physics 52: 479–487 (1988)
7. Licata, I. and Fiscaletti, D. Quantum Potential: Physics, Geometry and Algebra. Springer (2014)
8. Licata, I. and Fiscaletti, D. A Fisher-Bohm geometry for quantum information. EJTP 11: 71-88 (2014)
9. Wheeler, J.: It from bit. Proceedings 3rd International Symposium on Foundations of Quantum Mechanics, Tokyo (1989).
10. Chiribella, G., D'Ariano, G.M., Perinotti, P. Informational derivation of quantum theory. Physical Review A. 84, 012311 (2011)
11. Leifer, M.S. and Spekkens, R.W. Towards a formulation of quantum theory as a causally neutral theory of Bayesian inference. Physical Review A. 88, 052130 (2013)
12. Timpson, C. Quantum information theory and the foundations of quantum mechanics. Oxford: Oxford University Press (2010)
13. Vedral, V. Decoding reality: The universe as quantum information. Oxford University Press (2010)
14. Vachaspati, T., Stojkovic, D, Krauss, L. Observation of incipient black holes and the information loss problem. Phys. Rev. D 76, 24005 (2007)
15. Kak, S. The initialization problem in quantum computing. Foundations of Physics 29: 267-279 (1999)

16. Kak, S. A three-stage quantum cryptography protocol. Foundations of Physics Letters 19: 293-296 (2006).
17. Landauer, R. The physical nature of information. Phys. Lett. A 217, 188- 193 (1996)
18. Holevo, A.S. Bounds for the quantity of information transmitted by a quantum communication channel. Problems of Information Transmission 9: 177–183 (1973)
19. Holevo, A.S. Coding theorems for quantum channels. arXiv:quant-ph/9809023 (1973)
20. Penrose, R. The Road to Reality. Vintage Books (2004)
21. Kak, S.: Quantum information and entropy. International Journal of Theoretical Physics 46, 860-876 (2007)
22. Kak, S. Measurement complexity and oracle quantum computing. NeuroQuantology 12: 374-381 (2014)
23. Massar, S. and Popescu, S. Optimal extraction of information from finite quantum ensembles. Phys. Rev. Lett. 74 (1995)
24. Bagan, E., Baig, M., Muñoz-Tapia, R. Optimal scheme for estimating a pure qubit state via local measurements. Phys. Rev. Lett. 89 (2002)
25. Dicke, R.H. Interaction-free quantum measurements: A paradox? Am. J. Phys. 49: 925 (1981)
26. Elitzur, A.C.and Vaidman, L. Quantum mechanical interaction-free measurements. Foundations of Physics 23: 987-97 (1993)